%% file: deepfind.tex
\documentclass{new_tlp}

\newcommand{\appendixref}[1]{%
\ifcsname appendices\endcsname %
 \ref{#1}%
  \else %
  the supplementary material%
 \fi %
}

\newcommand{\appendixfigref}[2]{%
\ifcsname appendices\endcsname %
 \ref{#1}%
  \else %
  #2 in the supplementary material%
 \fi %
}

\newlength{\figwidth}
\newcommand{\ignore}[1]{}
\newcommand  {\bigsqcap}  {\mathop{\lower.1ex\hbox{\Large$\sqcap$}}}

\newcommand  {\implies} {\mbox{$\Rightarrow$}}

\def\tuple#1{\langle #1 \rangle}

\def\anno#1{{\ooalign{\hfil\raise.07ex\hbox{\small{\rm #1}}\hfil\crcr\mathhexbox20D}}}

\newcommand{\kbd}[1]{\mbox{\tt #1}}

\newcommand{\calls}{\ensuremath{\mathtt{calls}}\xspace}
\newcommand{\success}{\ensuremath{\mathtt{success}}\xspace}

\newcommand{\DD}{\mbox{$D_\alpha$}}

\newcommand*{\seq}[2][n]  {{#2_{1}, \allowbreak \ldots, \allowbreak #2_{#1}}}

\newcommand{\andtree}{generalized {\sc and} tree\xspace}
\newcommand{\andtrees}{generalized {\sc and} trees\xspace}
\newcommand*{\andnode}[1][L]{\ensuremath{\langle {#1},\theta_c,\theta_s \rangle}}

\newcommand{\csemantics}{\ensuremath{\llbracket P \rrbracket_Q}}

\newcommand{\callingcontext}{\ensuremath{calling\_context(L,P,Q)}}

\newcommand{\CP}{\mbox{\it CP}}

\def\tuple#1{\langle{#1}\rangle}

\newcommand{\A}{\mbox{$\cal A$}}
\newcommand{\Q}{\mbox{$\cal Q$}}

\usepackage{fancybox}
\usepackage{latexsym}
\usepackage{verbatim}
\usepackage{xcolor}
\usepackage{amssymb}
\usepackage{algorithm}
\usepackage{algpseudocode}
\usepackage[T1]{fontenc}
\usepackage{latexsym}
\usepackage{stmaryrd}
\usepackage{xspace}
\usepackage{multirow}
\usepackage{hyperref}
\usepackage{longtable}
\usepackage{tikz}
\usepackage{subcaption}

\usepackage{listings}
\newcommand{\prettylstformat}[0]{
\lstset{language=Prolog,
        numbers=left,numberstyle=\tiny,stepnumber=1,numbersep=8pt,
        frameround=fttt,
        frame=ltrb,
        showstringspaces=false,
        basicstyle=\scriptsize\ttfamily,
        commentstyle=\color{gray},
        breaklines=true,breakatwhitespace=true,
        showlines=true,
        showspaces=false,showtabs=false,
        keywords={pred,prop, true},
        escapeinside={~~}
      }}

\def\tuple#1{\langle{#1}\rangle}

\newenvironment{example-box}{\begin{example}\rm}{$\Box$\end{example}}

\newcommand{\sketch}[1]{
}

\newcounter{mnotei}
\input{macros}

\title[Semantic Code Browsing]
      {Semantic Code Browsing\ $^{\small \thanks{This research has received
          funding from the 
          EU FP7 agreement no 318337, ENTRA, Spanish MINECO
          TIN2012-39391 \emph{StrongSoft} and TIN2015-67522-C3-1-R
          \emph{TRACES} projects, and the Madrid M141047003
          \emph{N-GREENS} program.}}$}

\author{Isabel Garc\'{i}a \and Jose F. Morales \and
  Manuel V. Hermenegildo}

\author[Isabel Garc\'{\i}a-Contreras, Jos\'{e} F. Morales,
  and Manuel V. Hermenegildo]{
    \hspace*{-7mm}   ISABEL GARC\'{I}A-CONTRERAS$^{1}$
    ~~ JOS\'{E} F. MORALES$^{1}$
    ~~ MANUEL V. HERMENEGILDO$^{1,2}$
\ \\
\ \\
   $^1$IMDEA Software Institute 
   \email{\{isabel.garcia, josef.morales,
     manuel.hermenegildo\}@imdea.org} \\
   $^2$School of Computer Science, Technical University of Madrid
   (UPM) 
   \email{manuel.hermenegildo@upm.es}\\
}

\submitted{May 6, 2016}
\revised{July 8, 2016}
\accepted{July 22, 2016}

\begin{document}

\maketitle

\begin{abstract}
Programmers currently enjoy access to a very high number of code
repositories and libraries of ever increasing size.  The ensuing
potential for reuse is however hampered by the fact that searching
within all this code becomes an increasingly difficult task.
Most code search engines are based on syntactic techniques such as
signature matching or keyword extraction. However, these techniques
are inaccurate (because they basically rely on documentation) and at
the same time do not offer very expressive code query languages.
We propose a novel approach that focuses on querying for
\emph{semantic} characteristics of code obtained automatically from
the code itself.  Program units are pre-processed using static
analysis techniques, based on abstract interpretation, obtaining safe
semantic approximations.  A novel, assertion-based code query language
is used to express desired semantic characteristics of the code as
partial specifications.  Relevant code is found by comparing such
partial specifications with the inferred semantics for program
elements.
Our approach is fully automatic and does not rely on user annotations
or documentation. It is more powerful and flexible than signature
matching because it is parametric on the abstract domain and
properties, and does not require type definitions.  Also, it
reasons with relations between properties, such 
as implication and abstraction, rather than just equality.  It is also
more resilient to syntactic code differences.
We describe the approach and report on a prototype implementation
within the Ciao system.  
\end{abstract}

\begin{keywords}
Semantic Code Search, Abstract Interpretation, Assertions.
\vspace*{-3mm}
\end{keywords}

\vspace*{-2mm}
\section{Introduction}
\label{sec:intro}

The code sizes of current software systems and libraries grow
continuously. The open-source revolution implies that programmers
now enjoy access to many repositories which are very often
large.  While this abundance brings great potential for code
reuse, with the ensuing promise of coding time savings,
it also brings about a new problem: searching within these code bases 
is becoming an increasingly difficult task.
%
Most code search engines have so far addressed this problem through
syntactic techniques such as keyword extraction and signature
matching.
\cite{maarek1991information} is an early example of the work based on
information retrieval techniques.  It used keywords extracted from \texttt{man}
pages written in natural language.
More recent code search engines like Black Duck Open Hub
(\url{http://code.openhub.net}) use the same techniques but including
also keyword extraction from variable names in the code itself. They
combine those keywords with relatively simple characteristics of the kind of
code the user is looking for (e.g., whether it is classes, methods, or
interfaces).
Other recent work has used a similar approach combined with ranking
techniques. For example,~\cite{mcmillan2012recommending} use 
annotations in code instead of man pages in order to cluster features
from Java packages. They also incorporate the idea that multiple users
will rank over time how packages match searches.
Google code search (\url{https://github.com/google/codesearch}) is
based on regular expressions.
While keyword and regular expression search is obviously useful, the
fact that these techniques rely on documentation (including the names
of identifiers in the code) means that they also have shortcomings.
They are clearly of limited use if the code has no comments, existing
comments are wrong, they are written in a different (natural) language,
or other elements like variable, module, or
procedure names are not representative and/or not easy to match
against.

An alternative to keyword search is to query instead the signatures
present in code, an approach already proposed
in~\cite{rollins1991specifications} for finding code written in a
functional language.  The solver within $\lambda$Prolog was used to
\emph{match} the signatures in code against some pre- and
post-condition specifications used as search keys.  The Haskell code
browser, Hoogle~\cite{mitchell:hoogle_19_nov_2008}, combines signature
matching with keyword matching.  In the same
line~\cite{reiss2009semantics} combines these two techniques with test
cases as a means for specification.  Signature matching is a more
formal approach than keyword matching, but it is still essentially
syntactic, relies on the presence of signatures in the program, and is
limited to the properties of the language of the signatures, i.e.,
generally types.

We propose a new approach that focuses on querying for \emph{semantic}
characteristics of code that are inferred automatically from the code
itself.  Instead of relying on user-provided signatures, comments, or
identifier names, the code bases are pre-analyzed using static
analysis techniques based on abstract interpretation, obtaining safe
approximations of the semantics of the program.  The use of different
abstract domains allows generating a wide (and user extensible)
variety of properties (generalized types, instantiation modes,
variable sharing, constraints on values, etc.) that can be queried.
To this end we also propose a flexible code query language based on
assertions that expresses specifications composed of these very
general properties. These abstract query specifications are used to
reason against the abstract semantics inferred for the code, in order
to select code elements that comply with the queries.

Our approach is fully automatic and does not rely on user annotations
or documentation.  Although assertions in the code can also help the
analysis, they are not needed, i.e., the approach works even if the
code contains no assertions or signatures, since the program semantics
is inferred by the abstract interpreter. It is thus more powerful than
signature matching methods (which it subsumes), which require such
signatures and/or type definitions.
The proposed approach also reasons with relations between
properties, such as implication and abstraction, rather than just
matching, which allows much more expressive search and more accurate
results.  Our approach is also much more flexible, since it is
parametric on the abstract domain and properties, i.e., the inference
and the search can be based on any property for which an abstract
domain is available and not just syntactic match of the properties in
the signature language (generally types). It can also be tailored
through new abstract domains to fit particular applications.
Our approach can be more powerful than (and in any case is
complementary to) keyword-based information-retrieval systems because
its is based on a semantic analysis of the code, and is thus
independent of documentation. It is also more resilient to syntactic
differences (including code obfuscation techniques) such as, e.g., non
descriptive names of functions/variables.
Given their complementary nature, our implementation actually combines
the two approaches of semantic and keyword-based search. Since the
combination is straightforward, it is not described herein.

\vspace*{-5mm}
\section{Preliminaries, Abstract Interpretation, and Assertions}
\label{sec:prelim}

We denote by \VS, \FS, and \PS\ the set of variable, function, and
predicate symbols, respectively. Variables start with a capital
letter. Each $p \in \PS$ is associated with a natural number called
its \emph{arity}, written $\ar{p}$ or $\ar{f}$.  The set of terms
$\TS$ is inductively defined as follows:\footnote{We limit for
  simplicity the presentation to the Herbrand domain, but the approach
  and results apply to constraint domains as well. In the rest of the
  paper we will refer interchangeably to substitutions or constraints,
  and to the current substitution or the constraint store.}  $\VS
\subset \TS$, if $f \in \FS$ and $t_1, \ldots, t_n \in \TS$ then
$f(t_1,\ldots,t_n) \in \TS$ where $\ar{f}=n$.  An {\em atom} has the
form $p(t_1,...,t_n)$ where $p$ is a predicate symbol and $t_i$ are
terms. A {\em predicate descriptor} is an atom $p(\seq{X})$ where
$\seq{X}$ are distinct variables. A {\em clause} is of the form $H
\mbox{\tt :-} B_1,\dots ,B_n$ where $H$, the {\em head}, is an atom
and $B_1,\dots ,B_n$, the {\em body}, is a possibly empty finite
conjunction of atoms.
We assume that all clause heads are normalized,
i.e., $H$ is of the form of a predicate descriptor.
Furthermore, we require that each clause head of a predicate $p$ have
identical sequence of variables $X_{p_1},...,X_{p_n}$. We call this
the \emph{base form} of $p$.  This is not restrictive since programs
can always be put in this form, and it simplifies the presentation.
However,
in the examples and in the
implementation we handle non-normalized programs.
A {\em definite (constraint) logic program}, or {\em program}, is a
finite sequence of clauses. The concrete semantics used for reasoning
about goal-dependent compile-time semantics of logic programs will use
the notion of \andtrees%
~\cite{bruy91}.
A \andtree represents the execution of a query to a Prolog 
predicate.
Basically, every node of a \andtree contains a call to a predicate,
adorned on the left with the call substitution 
to that predicate, and 
on the right with the
corresponding success substitution.  The concrete semantics of a
program $P$ for a given set of queries $Q$, \csemantics, is the set of
\andtrees\ that represent the execution of the queries in $Q$ for the
program $P$.  We will denote a node in a \andtree with \andnode, where
$L$ is the call to a predicate $p$ in $P$, and $\theta_c, \theta_s$
are the call and success substitutions over $vars(L)$ adorning the node,
respectively.
The \callingcontext\ of a predicate given by the predicate descriptor
$L$ defined in $P$ for a set of queries $Q$ is the set $\{ \theta_c |
\exists T \in \csemantics\ s.t.\ \exists \andnode[L']\ in\ T \wedge
\exists\sigma\in ren\ L\sigma = L' \}$, where 
$ren$
is a set of renaming substitutions over variables in the
program.
We denote by $answers(P,Q)$ the set of answers (success substitutions)
computed by $P$ for query $Q$.

\vspace*{-3mm}
\paragraph{\textbf{Inferring the Program Semantics by Abstract interpretation}:}
\label{sec:absint}

As mentioned in the introduction, our approach for finding predicates
semantically is based on pre-processing program units using static
analysis techniques, in order to obtain safe approximations of the
semantics of the predicates in these units.  Our basic technique for
this purpose is {\em abstract interpretation} \cite{Cousot77-short}, 
an approach for static program analysis in which execution of the
program is simulated on an {\em abstract domain} ($D_\alpha$) which is
simpler than the actual, {\em concrete domain} ($D$). Although not
strictly required, we assume \DD~has a lattice structure with meet
$(\sqcap)$, join $(\sqcup)$, and less than $(\sqsubseteq)$
operators. Abstract values and sets of concrete values are related via
a pair of monotonic mappings $\langle \alpha, \gamma \rangle$: {\em
  abstraction} $\alpha: D\rightarrow D_\alpha$, and {\em
  concretization} $\gamma: D_\alpha\rightarrow D$. Concrete operations
on $D$ values are approximated by corresponding abstract operations
on \DD~values. The key result for abstract interpretation is that it
guarantees that the analysis terminates, provided that $D_\alpha$
meets some conditions (such as finite ascending chains) and that the
results are safe approximations of the concrete semantics (provided
\DD~safely approximates the concrete values and operations).

\vspace*{-3mm}
\paragraph{Goal-dependent abstract interpretation:}
While our approach is valid for any analysis, 
we will be using for concreteness 
goal-dependent abstract
interpretation,
in particular the PLAI algorithm~\cite{ai-jlp}, available within the
Ciao/CiaoPP
system~\cite{ciaopp-sas03-journal-scp,hermenegildo11:ciao-design-tplp-short}.
PLAI takes as input a program $P$, an abstract domain
$D_\alpha$,\footnote{Also, a set of abstract domains.}
and an abstract initial call pattern\footnote{We use sets of
  calls patterns in subsequent sections --the extension is straightforward.
}
$\Q_{\alpha} = L\mbox{:}\lambda$, where $L$ is an atom,
and~$\lambda$ is a restriction of the run-time bindings of $L$
expressed as an abstract substitution $\lambda \in D_\alpha$.  
The algorithm computes a set of triples
$analysis(P,L\mbox{:}\lambda,D_\alpha)$ $=$
$\{\tuple{L_1,\lambda_{1}^c,\lambda_{1}^s},$ $\ldots,$
$\tuple{L_n,\lambda_{n}^c,\lambda_{n}^s}\}$.  In each
$\tuple{L_i,\lambda_{i}^c,\lambda_{i}^s}$ triple, $L_i$ is an atom,
and $\lambda_{i}^c$ and $\lambda_{i}^s$ are, respectively, the
abstract call and success substitutions, elements of $D_\alpha$.
Let $Q$ be the set of concrete queries described by
$L\mbox{:}\lambda$, i.e., $Q = \{L\theta \mid\
\theta\in\gamma(\lambda)\}$.  In addition to termination, correctness
of abstract interpretation provides the following guarantees:
\begin{itemize}
\vspace*{-2mm}
\itemsep=1pt
\item The abstract call substitutions cover all the concrete calls
  which appear during execution of the initial queries in $Q$.
Formally, 
$\forall p'$ in $P$ $\forall \theta_c \in calling\_context(p',P,Q)$
$\exists \tuple{L',\lambda^c,\lambda^s}\in
analysis(P,L\mbox{:}\lambda)$ s.t.\ $\theta_c \in
\gamma(\lambda^c)$, where $L'$ is a base form of $p'$.
\item The abstract success substitutions cover all the concrete
  success substitutions which appear during execution, i.e., $\forall
  i=1\ldots n$ $\forall \theta_c\in\gamma(\lambda^c_i)$ (which, as we saw
  before, cover all the calling contexts)
  if $L_i\theta_c$ succeeds in $P$ with computed answer $\theta_s$
  then $\theta_s\in\gamma(\lambda_i^s)$.
\vspace*{-2mm}
\end{itemize}
The abstract interpretation process
is monotonic, in the sense that
more specific initial call patterns yield more precise 
analysis results.
As usual, $\bot$ denotes the abstract substitution such that
$\gamma(\bot)=\emptyset$. A tuple $\tuple{P_j,\lambda_j^c,\bot}$
indicates that all calls to predicate $p_j$ with substitution
$\theta\in\gamma(\lambda_j^c)$ either fail or loop, i.e., they do not
produce any success substitutions.

\vspace*{-3mm}
\paragraph{Multivariance:} The analysis (as well as the assertion
language presented later) is designed to discern among the various
usages of a predicate.  Thus, multiple usages of (types of calls to) a
procedure can result in multiple descriptions in the analysis
output, i.e., for a 
given predicate $P$ multiple $\tuple{P,\lambda^c,\lambda^s}$ triples
may be inferred and queried.  This will allow
finding code more accurately.  More precisely, the analysis is said to
be {\em multivariant on calls} if more than one triple
$\tuple{P,\lambda_{1}^c,\lambda_{1}^s}$, $\ldots,$
$\tuple{P,\lambda_{n}^c,\lambda_{n}^s}$ $n\geq 0$ with
$\lambda_i^c\neq\lambda_j^c$ for some $i,j$ may be computed for the
same predicate.
In this paper we use analyses that are multivariant on calls.

\vspace*{-3mm}
\paragraph{Analysis target:}

We will look for predicates in a predefined set of programs or
modules. Each of them will be analyzed independently and we will
denote with $analysis(m,\DD,\Q_{\alpha})$ the analysis of a module $m$
with respect to the set of call patterns $\Q_{\alpha}$ in domain \DD.
The reason for this kind of analysis is that normally users are
looking for independent libraries to reuse.
We assume for concreteness the Ciao module
system~\cite{ciao-modules-cl2000-short}. It is a strict module system,
i.e., a system in which modules can only communicate via their
interface. The interface of a module contains the names of the
exported predicates and the names of the imported modules. When
performing the analysis, only the exported predicates will be
considered for the initial calls.  We will use $exported(m)$ to
express the set of predicate names exported by module $m$.

An issue in the computation performed by $analysis(m,\DD,\Q_{\alpha})$
is that, from the point of view of analysis, the code of the module
$m$ to be analyzed taken in isolation is {\em incomplete}, in the
sense that the code for procedures imported from other modules is not
available to analysis.  The direct consequence is that, during the
analysis of a module $m$, there may be calls $P:\CP$ such that the
procedure $P$ is not defined in $m$ but instead it is imported from
another module $m'$. A number of alternatives are available (and
implemented in the system in which we conduct our experiments, Ciao)
in order to deal with these inter-modular
connections~\cite{mod-an-lopstrbook-shortalt}.  We assume, without
loss of generality, that for these external calls, we will trust the
assertions present in the imported modules for the predicates they
export, and use their information in the individual module analysis.
\vspace*{-2mm}

\paragraph{\textbf{Traditional Assertions:}}
\label{sec:assertions}

Assertions are linguistic constructions for expressing abstractions
of the meaning and behavior of programs.
Herein,
we will use for concreteness the \texttt{pred} assertions
of~\cite{assert-lang-disciplbook-short}
Such \texttt{pred} assertions allow 
stating sets of \emph{preconditions} and \emph{conditional
  postconditions} on the state (current substitution or constraint
store) that hold or must hold for a given predicate.
These assertions are instrumental for many purposes ranging from
expressing the results of analysis to providing partial specifications
which are then very useful for detecting deviations of behavior
(symptoms) with respect to such assertions, or to ensure that no such
deviations exist (correctness)~\cite{assert-lang-disciplbook-short}.
A \texttt{pred} assertion is of the form:
\vspace*{-2mm}
$$ \kbd{:- pred } Head \kbd{ : } Pre \kbd{ => } Post \kbd{.} $$

\vspace*{-1mm}
\noindent
where $Head$ is a normalized atom that denotes the predicate that the
assertion applies to, and the $Pre$ and $Post$ are 
conjunctions of ``\texttt{prop}'' atoms, i.e., of atoms whose
corresponding predicates are declared to be
\emph{properties}~~\cite{assert-lang-disciplbook-short,assrt-theoret-framework-lopstr99}. Both
$Pre$ and $Post$ can be empty conjunctions (meaning true), and in that
case they can be omitted.
The following example illustrates the basic concepts involved:

\begin{example-box}
  \label{ex:length}
  These assertions describe different modes for calling a \kbd{length}
  predicate: either for {\tt (1)} generating a list of length {\tt N},
  {\tt (2)} to obtain the length of a list {\tt L}, or {\tt (3)} to
  check the length of a list: 
\begin{scriptsize}
\prettylstformat
\begin{lstlisting}
:- pred length(L,N) : (var(L), int(N))  => list(L). %(1)
:- pred length(L,N) : (var(N), list(L)) => int(N).  %(2)
:- pred length(L,N) : (list(L), int(N)).            %(3)

:- prop list/1.    list([]).     list([_|T]) :- list(T). ~\label{line:list}~
\end{lstlisting}
\end{scriptsize}
Note also the definition of the \texttt{list/1} property (in this case
a regular type) in line~\ref{line:list}.  Other properties
(\texttt{int/1}, a base regular type, and \texttt{var/1}, a mode)
are assumed to be loaded from the libraries (\texttt{native\_props} in
Ciao for these properties).
\vspace*{-2mm}
\end{example-box}

The following definition relates a set of assertions for a predicate
to the nodes which correspond to that predicate in the \andtree for
the current program $P$ and initial set of queries $\Q$:
\begin{definition}[The Set of Assertion Conditions for a Predicate]
\label{def:assr_cond}
Given a predicate represented by a normalized atom $Head$, and a
corresponding set of assertions $\A = \{A_1 \ldots A_n\}$, with $A_i =
``\texttt{:- pred } Head \texttt{ : } Pre_i \texttt{ => } Post_i
\texttt{.}$'' the set of \emph{assertion conditions} for $Head$
determined by $\A$ is $\{ C_0, C_1, \ldots , C_n\}$, with:
  \vspace*{-1mm}
  \[
    C_i = \left\{
    \begin{array}{ll}
      \calls(Head,\bigvee _{j = 1}^{n} Pre_j)
    & ~~~~i = 0 
    \\
      \success(Head,Pre_i,Post_i)
    & ~~~~i = 1..n
    \end{array}
    \right.
  \]
\end{definition}
\vspace*{-1mm}
\noindent
where \calls(Head,Pre) states conditions on $\theta_c$ in all
nodes \andnode{} where $L \wedge Head$ holds, and
\success(Head,Pre,Post) refers to conditions on $\theta_s$ in all
nodes \andnode where $L \wedge Head$ and $Pre \wedge \theta_c$ hold.

\vspace*{1mm}
The assertion conditions for the assertions in the example above are: 
\begin{small}
\vspace*{-1mm}
  \[
   \left\{
     \begin{array}{lllll}
       calls( & length(L,N), & ((var(L) \wedge int(N)) \vee (var(N)
            \wedge list(L)) \vee (list(L) \wedge int(N)) ), \\ [1mm]
       success(& length(L,N), & (var(L) \wedge int(N)), \ \ \ list(L) ),  \\
       success(& length(L,N), & (var(N) \wedge int(L)), \ \ \ int(N)  ), \\
    \end{array}
    \right\}
  \]
\end{small}
\vspace*{-6mm}

\section{Abstract Code Search}
\label{sec:search}

In this section we propose the mechanism for defining abstract
searches for predicates.
Our objective now is not describing concrete predicates as before, but
rather to state some desired semantic characteristics and perform a
search over the set of predicates in some code $P$ (our set of
modules) looking for a subset of predicates meeting those
characteristics.  To this end we define the concept of~\emph{query
  assertions}, inspired by the \emph{anonymous assertions}
of~\cite{asrHO-ppdp2014}.  This requires extending our syntax so that
in the normalized atoms that appear in the $Head$ positions of these
assertions, the predicate symbol can be a variable from \VS.

\begin{definition}[Query assertion]
  A query assertion is an expression of the form: 
  \fbox{$
  \kbd{:- pred } L \kbd{ : } Pre \kbd{ => } Post \kbd{.}
  $} 
  where L is of the form $X(V_1, ..., V_n)$ and $Pre$ and $Post$ are
  (optional) DNF formulas of prop literals.
\end{definition}

We will use this concept to express conditions on the search.  The
intuition is that a query assertion is an assertion where the variable
$X \in VS$ in the predicate symbol location of $L$ will be
instantiated during the search for code to predicate symbols from $PS$
that comply with some query assertions.
The following predicate defines the search: 

\begin{definition}[Predicate query]
    \label{def:query}
    A predicate query is of the form: 
   \fbox{$ \kbd{?- findp(\{}\ As\ \kbd{\}, M:Pred/A, Residue, Status).} $}
    where:
    \begin{itemize}
   \vspace*{-2mm}
   \itemsep=1pt
   \item \textbf{\emph{As}} is a set of query assertions, with the
      same arity and the same variable \texttt{Pred} as main functor
      of the different assertion $Head$s.  This set can also include
      definitions of properties (e.g.,
      regtypes~\cite{gallagher-types-iclp94-short,eterms-sas02-short} or other \kbd{prop}erties) 
      used in the query assertions.
    \item \textbf{M:Pred/A} is a predicate descriptor, referring to a
      predicate \kbd{Pred} with arity \kbd{A} and defined in module
      \textbf{M} that corresponds to the information in the other
      arguments.
    \item \textbf{Residue} is a set of pairs of type
      $(condition, list(domain, status))$ which express the result of
      the proof of each condition in each domain. The status will be
      \emph{checked} for those conditions that were proved to hold in
      $domain$, \emph{false} if they were proved not to hold, and
      \emph{check} for conditions for which nothing could be proved.
    \item \textbf{Status} is the overall result of the proof for the
      whole set of conditions in the query assertion. It will be
      \emph{checked} if all conditions are proved to be checked.
      \emph{false} if one condition is false, and \emph{check} if
      neither \emph{checked} nor \emph{false} can be proved. If
      \textbf{Status} is instantiated to e.g., \emph{checked} in the
      query, only matching predicates are returned.
    \end{itemize}
\vspace*{-2mm}
\end{definition}

Predicate queries are our main means for conducting the semantic
search for predicates.  The query assertions and property definitions
in $As$ induce a series of $calls$ and $success$ assertion conditions
(as per Def.~\ref{def:assr_cond}) which are used to perform the
filtering of candidate predicates.  I.e., the \calls conditions encode
that the admissible calls of the matching predicates should be within
the set of $Pre$ conditions.
The \success conditions encode that, if $Pre$ holds at the time of
calling the matching predicate, and the execution succeeds, then the
$Post$ conditions hold.

\begin{example-box}
  Given code $P$, the predicate query: 
\prettylstformat
\lstset{numbers=none}
\begin{lstlisting}
?- findp({ :- pred X(A,B) : (list(A), var(B)) => int(B). }, M:X/2, Residue, checked).
\end{lstlisting}
  indicates that the user is looking for predicates $p \in P$ with
  $\ar{p} = 2$, which allow calls in which the first argument is instantiated
  to a list and the second is a free variable, and that, when called
  in this way, if $p$ succeeds, their second argument will be
  instantiated to an integer. A predicate that matches this query
  is, for example, the \kbd{length/2} predicate 
  of Ex.~\ref{ex:length},  which we assume defined in
  module \kbd{lists}. The call to \texttt{findp} would then unify
  \kbd{M:X} to \kbd{lists:length}.
  \kbd{Residue} would contain the explanation of why the predicate
  matches (all conditions would be checked in this case; these
  conditions are illustrated later in other examples).
  Other possible matching predicates would be returned via backtracking.
\end{example-box}

We now address how a predicate matches the conditions in a predicate
query in the form of Def.~\ref{def:query}. To this end we
provide some definitions (adapted from
\cite{spec-jlp,assrt-theoret-framework-lopstr99}) which will be
instrumental in order to connect the literals in query assertions to
the results of analysis.

\begin{definition}[Trivial Success Set of a Property Formula]
  \label{def:trivial-suc-set}
  Given a
  conjunction $L$ of properties and the definitions for each of these
  properties in $P$, we define the {\em trivial success set} of $L$ in
  $P$ as:
  \vspace*{-1mm}
  $$
    TS(L,P)= 
\{ \bar{\exists}_{L}\theta \ | \exists \theta'\in answers(P,(L,\theta)) \mbox{ s.t. }
\theta\models\theta' \}.
  $$
\end{definition} 
\vspace*{-1mm} where $\bar{\exists}_{L}\theta$ denotes the projection
of $\theta$ onto the variables of $L$.  Intuitively, it is the set of
constraints $\theta$ for which the literal $L\theta$ succeeds without
adding new ``relevant'' constraints to $\theta$ (i.e., without
constraining it further).

For example, given the following program $P$:
\prettylstformat
\begin{lstlisting}
list([]).
list([_|T]) :- list(T).
\end{lstlisting}
\vspace*{-1mm}
and $L=\texttt{list(X)}$, both $\theta_1 = \{ X = [1,2] \}$ and
$\theta_2 = \{ X = [1,A] \}$ are in the trivial success set of $L$ in
$P$, but $\theta = \{ X = [1|\_] \}$ is not, since a call to
\texttt{(X = [1|\_], list(X))} will instantiate the second argument of
$[1|\_]$. We now define abstract counterparts for
Def.~\ref{def:trivial-suc-set}:

\begin{definition}[Abstract Trivial Success Subset of a Property Formula]
  \label{def:abstract-trivial-suc-set}
  Given a conjunction $L$ of properties, the definitions for each of
  these properties in $P$, and an abstract domain $D_\alpha$, an
  abstract constraint or substitution
  $\lambda^-_{TS(L,P)} \in D_\alpha$ is an {\em abstract trivial
    success subset} of $L$ in $P$ iff
  $\gamma(\lambda^-_{TS(L,P)})\subseteq TS(L,P)$.
\end{definition}

\vspace*{-2mm}
\begin{definition}[Abstract Trivial Success Superset of a Property Formula]
  Under the same conditions of
  Def.~\ref{def:abstract-trivial-suc-set} above, an abstract constraint or
  substitution $\lambda^+_{TS(L,P)}$ is an {\em abstract trivial
    success superset} of $L$ in $P$ iff
  $\gamma(\lambda^+_{TS(L,P)})\supseteq TS(L,P)$.
\end{definition}

\vspace*{-1mm}
\noindent
I.e., $\lambda^-_{TS(L,P)}$ and $\lambda^+_{TS(L,P)}$ are, respectively, a
safe under-approximation and a safe over-approximation of the trivial
success set for the property formula $L$ with definitions $P$.

\medskip 

We assume that the code $P$ under consideration has been analyzed for
an abstract domain $D_\alpha$, for a set of queries \Q. Let
$\Q_\alpha$ be the representation of those queries, i.e., it is the
minimal element of $D_\alpha$ so that $\gamma(\Q_\alpha)\supseteq \Q$.
We derive $\Q_\alpha$ from the code by including in it queries for all
exported predicates, affected by the calls conditions of any
assertions that appear in the code itself affecting such predicates
(this is safe because if analysis is not able to prove them, they will
be checked in any case via run-time checks).
If no assertions appear in the code for a given exported predicate,
the analyzer will assume $\top$ for the corresponding query.

We now relate, using the concepts above, the abstract semantics
inferred by analysis for this set of queries with the search process.
As stated in Def.~\ref{def:assr_cond}, a set of
assertions denotes different types of conditions (calls and
success). We provide the definitions for each type.

\vspace*{-1mm}
\begin{definition}[Checked Predicate Matches for a `calls' Condition]
\label{def:ch-call-ass}
A calls condition $\calls(X(V_1,\ldots, V_n),Pre)$ is abstractly
`checked' for a predicate $p \in P$ w.r.t.\ $Q_{\alpha}$ in $D_\alpha$
iff $\forall \tuple{L,\lambda^c,\lambda^s} \in analysis(P,\DD,
\Q_{\alpha})\ s.t.\ \exists\sigma\in ren,\ L=p(V'_1,\ldots,V'_n) =
X(V_1, \ldots, V_n)\sigma, \lambda^c \sqsubseteq
\lambda^-_{TS(Pre\ \sigma,P)}$. 
\end{definition}

\vspace*{-2mm}
\begin{definition}[False Predicate Matches for a `calls' Condition]
  A calls condition $\calls(X(V_1, \ldots, V_n),Pre)$ is abstractly
  `false' for a predicate $p \in P$ w.r.t. $Q_{\alpha}$ in $D_\alpha$
  iff $\forall \tuple{L,\lambda^c,\lambda^s} \in analysis(P,\DD,
  \Q_{\alpha})\ s.t.\ \exists\sigma\in ren,\ L=p(V'_1,\ldots,V'_n) =
  X(V_1, \ldots, V_n)\sigma, \lambda^c \sqcap \lambda^+_{TS(Pre\
    \sigma,P)} = \bot$.
\end{definition}

Note that in these definitions we do not use directly the $Pre$ and
$Post$ conditions, although they already are abstract
substitutions. This is because the properties in the conditions stated
by the user in assertions might not exist as such in $D_\alpha$.
However, it is possible to compute safe approximations
($\lambda^-_{TS(Pre,P)}$ and $\lambda^+_{TS(Pre,P)}$) by running the
analysis on the code of the property definitions using $D_\alpha$ (or
using the available trust assertions, for built-ins).  The fact that
the resulting approximations are safe ensures correctness of the
procedure both when checking calls and success conditions.
\begin{figure}
  \centering
  \prettylstformat
  \begin{lstlisting}
:- module(_, [my_length/2, get_length/2, check_length/2, gen_list/2], [assertions]).

:- pred my_length(L,N) : (list(L), var(N))  => int(N).
:- pred my_length(L,N) : (list(L), int(N)). 
~\color{gray}{:- true pred my\_length(L,N) : ( mshare([[L],[L,N],[N]]), var(N)). } ~
~\color{gray}{:- true pred my\_length(L,N) : ( mshare(L), ground(N) ).}            ~
my_length(L,N) :- length(L,N).

:- pred check_length(L,N) : (list(L), int(N)).
~\color{gray}{:- true pred check\_length(L,N) : (mshare(L), ground([N])). }    ~
check_length(L,N) :- length(L,N).

:- pred get_length(L,N) : (list(L), var(N)).
~\color{gray}{:- true pred get\_length(L,N) : (mshare([[L],[L,N],[N]]), var(N). } ~
get_length(L,N) :- length(L,N). 

:- pred gen_list(L,N) : (var(L), var(N)) => (list(L), int(N))
# "Generates a list of random elements of random size".
~\color{gray}{:- true pred gen\_list(L,N) : (mshare([[L],[L,N],[N]]), var(L), var(N)).} ~
gen_list(L,N) :- length(L,N).

% Implementation of length/2 ...
\end{lstlisting}

\vspace*{-3mm}
  \caption{Program with assertions stating different calls and
    (partial) analyzer output.}
  \label{fig:calls_program}
\vspace*{-2mm}
\end{figure}
\begin{example}\textit{Several checks against a `calls' condition.}
\label{ex:checked_calls}
Consider the program in Fig.~\ref{fig:calls_program} and the classic
sharing and freeness (\kbd{shfr}) abstract domain~\cite{freeness-iclp91}. 
Concentrating for now on calls only, this analysis will infer the
calls abstract states that are shown also in
Fig.~\ref{fig:calls_program}, as ``true'' \texttt{pred} assertions.
There, \kbd{var/1} and \kbd{ground/1} have the usual meaning and
\kbd{mshare/1} describes \emph{variable sharing} (intuitively, two
variables are in the same list if they may share, singletons mean that
there may also be other non-shared variables).  Note that, while the
\kbd{var/1} property is understood natively by the \kbd{shfr}
analyzer, other properties that appear in the assertions
(\kbd{list/1}, \kbd{int/1}, etc.) are not. However, they imply
groundness and freeness information.  The analysis approximates this
information to the \kbd{shfr} domain. In the case of built-ins such as
\kbd{int/1} this is done using the associated assertions in the
libraries.  Thus, if an argument is stated to have the property
\texttt{integer} on calls (i.e., it is bound to an integer at call
time, as in the second case of \kbd{my\_length} and
\kbd{check\_length}) it is expressed as a ground term in the
\kbd{shfr} domain.  In the case of properties that are defined by
programs, such as \kbd{list/1}, the property definition itself is
analyzed with the target domain (\kbd{shfr}). However, \kbd{shfr}
cannot infer too much about \kbd{list/1} since it does not have a
representation for ``definitely non-var.'' Other modes domains may be
able to infer ``non-var but not necessarily ground.''

Assume now that we would like to find predicates that generate tuples
of lists and their size, i.e., the predicate has to accept a usage in
which both of the arguments are free variables. This search can be
expressed with the following predicate query:
\vspace*{-1mm}
\prettylstformat
\lstset{numbers=none}
\begin{lstlisting}
?- findp({:- pred P(L, Size) : (var(L), var(Size)).}, M:P/A, Residue, Status).
\end{lstlisting}
\vspace*{-2mm}
The corresponding calls condition is: $\calls(X(L,Size),(var(L), var(Size)))$.
We discuss some interesting aspects of the search results: 

\begin{itemize}
  \vspace*{-2mm}
  \itemsep=0pt

\item \kbd{gen\_list/2}: This is obviously a predicate of interest in
  the context of the predicate query because it expects both of its
  arguments to be variables (plus, they will be bound during the
  execution to what we might want --a list and an integer). Formally,
  the conditions are proved to hold for this predicate, because: 
  \vspace*{-2mm}
 $$(\lambda_{TS((var(L), var(Size)),P)}^-= \{var(L), var(Size)\})
  \sqsupseteq (\lambda^c = {var(L), var(Size)}).$$

\item \kbd{check\_length/2}: This is not a predicate of
  interest because its calling modes require both arguments to be
  instantiated. Formally, the condition is abstractly false for
  \kbd{check\_length} because: 
  \vspace*{-2mm}
  $$(\lambda_{TS((var(L),
    var(Size)),P)}^+ = \{var(L), var(Size)\}) \sqcap (\{mshare(L),
  ground(Size)\} = \bot).$$
  
\item Both \kbd{my\_length/2} and \kbd{get\_length/2} are predicates
  which do not match what we are looking for, because they require at
  least one argument to be instantiated. However, using only the
  \kbd{shfr} domain this cannot be proved (it would if the domain
  could represent \kbd{nonvar/1}, which would then be incompatible
  with \kbd{var/1}).  The status for this condition for these
  predicates will be \emph{check}, meaning that (using the
  \texttt{shfr} domain only) the finder could not infer information
  regarding those conditions for the predicate, but still the user
  might be interested in it. \ \ $\Box$
  \end{itemize}
\end{example}
  
The point of filtering by calling modes is to avoid mixing behaviors.
This can be interesting for example with predicates that, depending on
the call, on success return in an argument either a free variable or
an instantiated term.  Consider an (admittedly not very nice)
predicate \kbd{read\_line(Line, Size)} such that if a line is
correctly read, its size will be \kbd{Size} and if not, \kbd{Size}
will be a free variable.
Assume that we would like instead an error to be displayed if the line
is not correctly read. Then, we need a predicate that requires
\kbd{Size} to be an integer. \kbd{check\_length} is a relevant
predicate then (and can be combined with \kbd{read\_line/2} as:
\kbd{read\_line(Line, Size), check\_length(Line, Size).}). In this case
\kbd{my\_length} is not useful, since it accepts the second
argument as a free variable.

\medskip Similarly to what we did for \calls conditions, we provide
definitions for stating whether a predicate matches for a given
\success condition and when it does not:

\vspace*{-1mm}
\begin{definition}[Checked Predicate Matches for a `success' Condition]
\label{def:match_pred_succ}
A success condition
$\success(X(V_1,\ldots, V_n),Pre,Post)$ is abstractly `checked'
for predicate $p \in P$ 
w.r.t.\ $Q_{\alpha}$ in $D_\alpha$ iff
$\forall \tuple{L,\lambda^c,\lambda^s} \in
analysis(P,Q_{\alpha})\ s.t.\ \exists\sigma\in ren,\ 
L=p(V'_1,\ldots,V'_n) =
X(V_1, \ldots, V_n)\sigma,
\lambda^c \sqsupseteq \lambda^+_{TS(Pre\ \sigma,P)} \rightarrow \lambda^s
\sqsubseteq \lambda^-_{TS(Post\ \sigma,P)}$.
\end{definition}

\vspace*{-2mm}
\begin{definition}[False Predicate Matches for a `success' Condition]
\label{def:false_pred_succ}
  A success condition
  $\success(X(V_1, \ldots, V_n),Pre,Post)$ is 
  abstractly `false' for $p \in P$ 
  w.r.t.\ $Q_{\alpha}$ in $D_\alpha$ iff
  $\forall \tuple{L,\lambda^c,\lambda^s} \in
  analysis(P,Q_{\alpha})\ s.t.\ \exists\sigma\in ren,\ 
  L=p(V'_1,\ldots,V'_n) = X(V_1, \ldots, V_n)\sigma,
  \lambda^c \sqsubseteq \lambda^-_{TS(Pre\ \sigma,P)} \wedge (\lambda^s \sqcap
  \lambda^+_{TS(Post\ \sigma,P)} = \bot)$.
\end{definition}

\vspace*{-3mm}
\begin{example-box}
  \textit{Several checks against a `success' condition.}
  \label{ex:checked-succ}
  Assume that we analyze the module in Fig.~\ref{fig:simple_program}
  with a shape abstract domain $D_\alpha$ ---in particular
  \texttt{eterms}~\cite{eterms-sas02-short} (regular types).
  Originally, the code had no assertions, so the analysis was
  performed for any possible entry. As before, the inferred
  information is provided by the analyzer as ``true'' \texttt{pred}
  assertions (we omit the calls conditions for simplicity). The
  relation among these inferred abstract elements is shown in lattice
  form in Fig.~\ref{fig:lattice}.
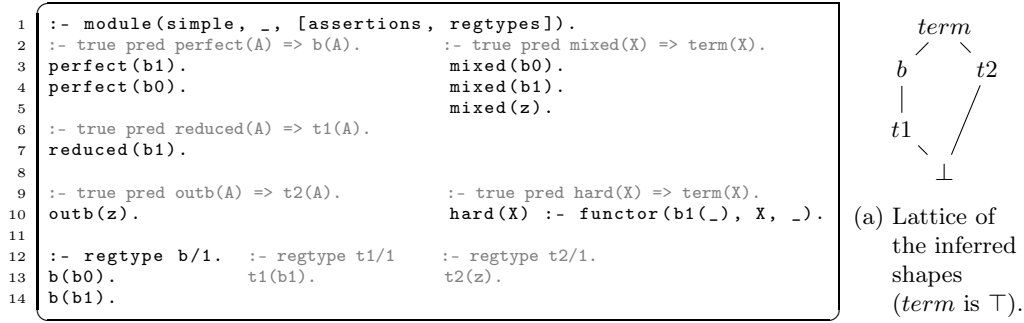
\begin{figure}
  \begin{subfigure}{0.77\textwidth}
    \prettylstformat
    \begin{lstlisting}
:- module(simple, _, [assertions, regtypes]).
~\color{gray}{:- true pred perfect(A) => b(A).} ~      ~\color{gray}{:- true pred mixed(X) => term(X). }~
perfect(b1).                      mixed(b0).
perfect(b0).                      mixed(b1).
                                  mixed(z).
~\color{gray}{:- true pred reduced(A) => t1(A).} ~
reduced(b1).                      
                                  
~\color{gray}{:- true pred outb(A) => t2(A).} ~        ~\color{gray}{:- true pred hard(X) => term(X).} ~
outb(z).                          hard(X) :- functor(b1(_), X, _).  

:- regtype b/1.  ~\color{gray}{:- regtype t1/1} ~   ~\color{gray}{:- regtype t2/1.} ~
b(b0).           ~\color{gray}{t1(b1).} ~          ~\color{gray}{t2(z).} ~ 
b(b1).
    \end{lstlisting}
  \end{subfigure}
  \hspace{+1mm}
  \begin{subfigure}{0.18\textwidth}
    \centering
    \begin{tikzpicture}[node distance=0.8cm]
      \title{Analysis lattice for \ref{fig:simple_program}}
      \node(top)                           {$term$};
      \node(t2)       [below right of=top] {$t2$};
      \node(b)      [below left of=top]  {$b$};
      \node(t1)      [below of=b]       {$t1$};
      \node(bot)            [below right of=t1]     {$\bot$};
      \draw(top)       -- (t2);
      \draw(top)       -- (b);
      \draw(b)      -- (t1);
      \draw(t1)      --  (bot);
      \draw(t2)      --  (bot);
    \end{tikzpicture}
    \captionsetup{format=hang}
    \caption{Lattice of\\ the inferred\\ shapes\\ ($term$ is $\top$).}
    \label{fig:lattice}
  \end{subfigure}
  \vspace*{-4mm}
  \caption{Another simple program with analysis information on success conditions.} 
  \label{fig:simple_program}
  \vspace*{-3mm}
\end{figure}
  The regular type $b$ was included in the program and $t1$ and
  $t2$ were inferred by the analyzer.
  Suppose that we execute the query: \hfill
  \ovalbox{\footnotesize\kbd{?- findp(\{:- pred P(V) : term(V) => b(V).\}, M:P/A, Residue, St).}}\\
  The \success condition of this query is $C = \success(X(V)$,
  term$(V)$, $b(V)$). We discuss how the predicates match this condition:
\vspace*{-2mm}
\begin{itemize}
  \item \kbd{perfect/1}. This predicate behaves
    exactly as specified in the predicate query, because on success it
    produces an output of the same type as specified.  Formally, the
    analysis infers $\tuple{perfect(V), \top(V), b(V)}$ and
    $\lambda^-_{TS(b,P)} = b$ (trivially). Then, $\lambda^s
    \sqsubseteq \lambda^-_{TS(b,P)}$, because $b\sqsubseteq b$.
  \item \kbd{reduced/1}. Intuitively, this predicate does not
    match as well as \kbd{perfect} but all possible outputs are within
    $\gamma(b)$, therefore, it is a valid predicate. Formally, the
    analysis infers $\tuple{reduced(V), \top(V), t1(V)}$, and
    $\lambda^-_{TS(b,P)} = b$ (trivially). As $t1 \sqsubseteq b$,
    i.e., $t1\ \implies\ b$, this predicate meets the condition of
    Def.~\ref{def:match_pred_succ} to be checked.
  \item \kbd{outb/1}. This predicate is of no use,
    because its output ($z$) is completely different (disjoint)
    from that specified in the query ($b$).  Formally, the analysis infers 
    $\tuple{outb(V), \top(V), t2(V)}$ and $\lambda^-_{TS(b,P)} =
    b$ so the conditions of the definition hold: $\lambda^c
    \sqsubseteq \lambda^-_{TS(Pre,P)}$ holds because
    $(\lambda^c = \top)\sqsubseteq (\lambda^-_{TS(term,P)} = \top)$
    and $(\lambda^s \sqcap \lambda^+_{TS(Post,P)} = \bot)$ holds because
    $(\lambda^s = t2) \sqcap ( \lambda^+_{TS(b,P)} = b) = \bot$.
  \end{itemize}
  Finally, we again have predicates
  (\kbd{mixed/1} and \kbd{hard/1})
  that are not \texttt{checked} or \texttt{false}.  As
  discussed before, this can be due to two reasons. The
  first is that the predicate may actually behave in such a way that
  the conditions in the query are really not checked or false. The
  second one is that the abstract domain may not provide accurate
  enough information to prove whether the conditions hold or not. In
  the case of 
  \label{ex:check}
  predicate \kbd{mixed/1}, it is the former:
  it is not what we are
  looking for because, although its possible outputs can be of type
  $b$, it can also produce type $t2$.
  Formally, the condition
  cannot be proved to hold or not, since the analysis inferred
  $\tuple{mixed(V), \top(V), \top(V)}$:

\vspace*{-2mm}
\begin{itemize}
\item It cannot be checked, because the output type is more general
  than specified, and therefore it does not satisfy the condition in
  Def.~\ref{def:match_pred_succ}:
  $(\lambda^c = \top) \sqsupseteq (\lambda^+_{TS(Pre,P)}) \rightarrow
  (\lambda^s = \top) \sqsubseteq (\lambda^-_{TS(b,P)} = b)$
  (true  $\rightarrow$ false).

\item It is also not false because some of the outputs are the ones
  required in the specification. Formally, it does not satisfy the
  second condition of Def.~\ref{def:false_pred_succ}:
  $(\lambda^s = \top) \sqcap (\lambda^+_{TS(b,P)}=b) = b \neq \bot$.
  \end{itemize}

\vspace*{-1mm}  
\noindent
  Predicate \kbd{hard/1} illustrates the latter case: 
  that an abstract domain may not be precise enough to find
  all matching predicates.
  Intuitively, the success condition of the example should hold
  because its output shape is more restrictive than
  specified. However, the analyzer cannot infer that its output will
  be always \kbd{b1} because \kbd{functor/3} can produce any atom, and
  thus the inferred tuple will be $\tuple{hard(V), \top(V),
    \top(V)}$. The reasoning to set the status of proof of this
  condition as check is the same as with \kbd{mixed/1}.
\vspace*{-3mm}
\end{example-box}

\paragraph{\textbf{Combining information from different domains:}}

Sometimes the information inferred using an abstract domain is not
accurate enough to prove whether a condition holds or not but the
information in another domain is. It depends on how the user expresses
the query, and how accurately the abstract properties of the query can
be approximated in each domain.  For example, in \kbd{:- pred X(A,B) :
  (list(A), var(B))}, the property \kbd{var(X)} cannot be represented
in the (standard) regular types domain (\texttt{eterms}), so it will
assume $\top$ for \kbd{B} which will lead to not being able to check
it.

Combining domains is a useful technique to increase accuracy.  An
assertion condition is proved to hold (status \texttt{checked}) or not
(status \texttt{false}) if the result can be proved in any analysis
domain. The reason for this is the correctness of the analysis, which
always computes safe approximations.  This ensures that properties
proved in each domain separately for the same set of queries cannot be
contradictory. At most, if a property can be proved in a domain, other
domains may not be accurate enough to decide that the property holds.
Summarizing, the status of a condition given its proof status for a
set of domains will be:
\vspace*{-5mm}
  \[   Status = \left\{
\begin{array}{ll}
      false & \mathrm{if\ proved\ false\ in\ at\ least\ one\ domain}\\
      checked & \mathrm{if\ proved\ checked\ in\ at\ least\ one\ domain}\\
      check & \mathrm{otherwise}
\end{array} 
\right.
\]

\vspace*{-7mm}
\begin{example}
  Assume the program in Fig.~\ref{fig:calls_program} and the analysis
  in Ex.~\ref{ex:checked_calls}, but that the \texttt{eterms} shape
  analysis is also performed:

\vspace*{1mm}
\begin{small}
  \begin{tabular}{ccc}      
    Predicate & $\lambda^c$ (eterms) &  $\lambda^c$ (shfr) \\ [-2mm]
    \hline
    $gen\_list(L,N)$ & $(term(L), term(N))$ & $(mshare([[L],[L,N],[N]]), var(L), var(N))$ \\
    $get\_length(L,N)$ & $(list(L), term(N))$ &  $(mshare([[L],[L,N],[N]]), var(N))$  \\
    $check\_length(L,N)$ & $(list(L), int(N))$ &$(mshare(L), ground([N]))$\\
    $my\_length(L,N)$ & $(list(L), term(N))$ & $(mshare(L), ground(N))$ \\
    $my\_length(L,N)$ & $(list(L), int(N))$ & $(mshare([[L],[L,N],[N]]), var(N))$
  \end{tabular}
\end{small}

The combination of both domains is really useful for proving certain
conditions because they complement each other.
Assume that we want to find a predicate that checks the length of a
list. The condition to be satisfied is $\calls(X(L, Size), (list(L),
num(Size)))$.  According to the definitions of matching, the results
in each domain will be:

\vspace*{1mm}
\begin{small}
  \begin{tabular}{c|ccc}
    PredName/A & eterms proof &  shfr proof & combined proof (Sum) \\ [-1mm]
    \hline\\ [-6mm]
    gen\_list/2 &  check & false & false \\
    get\_length/2 & check & false & false \\
    check\_length/2 & checked & check & checked \\
    my\_length/2 & check & check & check
\end{tabular}
\end{small}

The intuitive explanation of these results is:
\begin{itemize}
\vspace*{-3mm}
\itemsep=0pt
\item \kbd{gen\_list/2}: In the \kbd{eterms} domain this condition cannot
  be proved because the domain has no information about 
  \kbd{var}. However, in the \kbd{shfr} domain 
  it can be proved that the condition does not hold
  because it requires both arguments to be non-free
  variables, and the calling mode does the opposite. Then, that
  condition is false for this predicate.
\item \kbd{get\_length/2}: This case is similar to \kbd{gen\_list/2}: It
  cannot be proved in the types domain because one argument was
  specified with instantiation information but it can be proved in the
  modes domain that it is false.
\item \kbd{check\_length/2}: 
  Matches the condition in
  the \kbd{eterms} domain, because the shapes are exactly the ones we were
  looking for. For this predicate, the \kbd{shfr} domain is not
  necessary.
\item \kbd{my\_length/2}: At first sight this predicate matches the
  query because there is a calling mode that matches exactly as
  stated in the condition. However, according to the definition of
  calls condition, all admissible calling modes must be within the
  condition, and there is one calling mode that does not comply:
  the mode for calculating the length of the list. \ \ $\Box$
\vspace*{-3mm}
\end{itemize}

\vspace*{-2mm}
\end{example}

\vspace{-3mm}
\section{Prototype and evaluation}
\label{sec:evaluation}

We have developed and evaluated a prototype implementation on top of
the Ciao/CiaoPP system.
The system implements both the pre-analysis of the code base
and the user-level predicate matching search facilities,
against the analysis results.
As mentioned in Section~\ref{sec:absint}, by default modules are
analyzed individually and the analysis trusts the assertions for
imported predicates and the calls for exported predicates. However,
modular analysis can also be used, as discussed later. The analysis
results are cached on disk (as CiaoPP \emph{dump} files) and reused
while searching.
Each time the search is performed in a module, its corresponding
analysis dump is restored or it is reanalyzed with the abstractions
of the constraints in the query, and conditions are checked.
The algorithms that implement condition checking are
described in~\appendixref{sec:algomatch}.%
\ifcsname appendices\endcsname %
\else %
\footnote{We refer to the supplementary material 
  provided on-line for this paper at the TPLP archives.}
\fi %
%


\vspace*{-3mm}
\paragraph{\textbf{Searching with the prototype.}}
To demonstrate some of the potential of our approach,
consider looking in the Ciao libraries for code that operates with
graphs. First, we need to guess how graphs may be represented, i.e.,
their shape. 
Two possible guesses are:

\begin{lstlisting}
:- regtype math_graph(Graph).              :- regtype al_graph(_).
math_graph(graph(Vertices,Edges)):-        al_graph(A) :- list(A,al_graph_elem).
    list(Vertices), list(Edges, pair).     
                                           :- regtype al_graph_elem/1.   
:- regtype pair/1.                         al_graph_elem(Vertex-Neighbors) :-
pair((_,_)).                                   list(Neighbors).  
\end{lstlisting}

\vspace*{-1mm}
\noindent
where \kbd{math\_graph} is based on the mathematical definition: an
ordered pair (V, E) comprising a set V of vertices, together with a
set E of edges, which are 2-element subsets of V. The \kbd{al\_graph}
property captures an alternative adjacency list graph representation,
as a list of vertices and their corresponding neighbors.
A query assertion for finding code that uses the first representation
could be
\kbd{:- pred P(X,Y)} \kbd{=> math\_graph(Y)}.%
\footnote{As mentioned before, the user-defined shapes (or any other
  properties), in this case the regtypes above, must be included
  within the predicate queries. However, we just show the query
  assertion for brevity.}
The prototype finds \kbd{complete\_graph/2} and \kbd{cycle\_graph/2}
in module \kbd{named\_graphs.pl} (see
Fig.~\appendixfigref{fig:named_graphs}{A 1}) by matching this query
against the analysis results for the module. Note that this code is
found although this \kbd{named\_graphs.pl} module has \emph{no
  assertions or shape/regtype definitions}, i.e., it only contains
plain Prolog code.
Searching for the second representation, assume we look for code for
modifying a graph, i.e., that takes as input a graph and a list of
elements and produces a new
graph:\\
\kbd{:- pred P(A,B,C) : (al\_graph(A), list(B), var(C)) }\kbd{=> al\_graph(C)}. 
I.e.:

$C_1 = \calls(P(A,B,C), (al\_graph(A),list(B),var(C)))$ and

$C_2 = \success(P(A,B,C), (al\_graph(A),list(B),var(C)), al\_graph(C))$,

\noindent
No code is found for which both conditions hold, because \calls can be
checked only if the code has assertions (hand-written or inferred
modularly).  Therefore, we focus on finding predicates for which $C_2$
holds.
Since the conditions on the calls substitution
are very specific, we assume they were 
not considered by the default pre-analysis. 
We can refine the predicate matching by reanalyzing the predicates
starting from the calls values in the success conditions.
To ensure greater precision, we perform 
inter-modular analysis.
Under these conditions the prototype 
finds that in \kbd{add\_vertices/3},
\kbd{del\_} \kbd{vertices/3}, \kbd{add\_edges/3}, and
\kbd{del\_edges/3} the \success condition does hold (see
Fig.~\appendixfigref{fig:ugraphs}{A 2}).

\input statistics_checking

\newcommand{\NumMods}{63}

\vspace*{-3mm}
\paragraph{\textbf{Performance results.}}
To measure the effectiveness and performance of the approach, we have
set up an experiment that consists in analyzing 
part of the Ciao libraries and finding matching predicates of arity 1
to 4 for several assertion conditions.
The experiments were run on a Linux server (Intel Xeon CPU E7450,
2.40GHz) with 16GB of RAM.
As in the previous examples, we used the \texttt{shfr} and
\texttt{eterms} domains (the Ciao system includes however a large
number of other domains than can also be used in this application).
We selected~\NumMods{} modules from the Ciao libraries
all of which can be analyzed within 
1 minute for these abstract domains.
The detailed analysis statistics can be found 
in~\appendixref{sec:anatables}. The selection includes modules that
are relatively costly 
for the analyses and others where analysis is trivial (e.g.,
non-analyzable foreign code with trusted assertions) but useful for
the search.
The pre-analysis of all the modules took $45s$ ($660ms$ on
average), and the analysis dumps required $3.5MB$ of disk space 
($55.5KB$ on average).
Restoring the analysis results (for the~\NumMods{} modules) takes
$21.5s$ ($343ms$ on average). In the experiments this was done 
for each query, but 
note that since the size of the cached analysis is small it
can be kept in memory for subsequent queries.
The performance of matching, once the analysis results are available,
depends on the arity, the number of predicates available with that arity, and
the conditions specified in the query.  
Summarized timing results
are shown in Table~\ref{tab:ass_check_time}.  Columns represent the
number of assertion conditions in  each predicate query and rows
their arity (the parentheses show the number of predicates present in
the code with that 
arity). Cells represent the execution time needed to exhaustively
check the predicates in the~\NumMods{} modules. The \emph{(AVG)}
columns represent the average time per predicate: 
from $224\mu s$ (1 condition, 1 argument) to $130 ms$ (4
conditions, 3 arguments).
Summarizing, it takes on average $25s$ to execute a query, looking
in all~\NumMods{} modules, most of which ($21.5s$) is spent 
loading the pre-analysis.

\vspace*{-5mm}
\section{Conclusions}
\label{sec:conclusions}

We have proposed a novel approach to the code search problem based on
querying for \emph{semantic} characteristics of the programs against a
safe approximation of its semantics obtained via analysis.  We have
also discussed the advantages of our proposal over other approaches
such as keyword search or signature matching.
We have provided evidence that both the analysis and the search are
sufficiently efficient, despite the relatively naive implementation,
for practical use. 
Our implementation actually combines semantic code search  with
keyword-based and other types of search.
A number of other extensions are also in progress, such as allowing
permutations or extra arguments, and applying other program
transformations.
We believe the proposed approach has a number of additional
applications, such as, for example, detection of duplicated code.
While prototyped within the Ciao system, the techniques proposed,
based on abstract interpretation theory, are general and directly
applicable to other languages.

\vspace*{-4mm}
\begin{small}
\bibliographystyle{acmtrans}
\bibliography{clip,general}
\end{small}

\ifcsname appendices\endcsname

\appendix
\clearpage 

\vspace*{-20mm}

\centerline{{\LARGE \textbf{Appendices}\footnote{ In the version of
      this paper published in TPLP these appendices constitute the
      supplementary, on-line material associated with the paper. }}}
\ \\


\input{deepfind_appendix_common}

\fi
\label{lastpage}
\end{document}

%% file: macros.tex
\newcommand{\longPaperVersion}{true}
\newcommand{\shortlong}[2]
    {\ifthenelse{\equal{\longPaperVersion}{true}}{#2}{#1}}

\newcommand{\oneColFormat}{false}
\newcommand{\onecoltwocol}[2]
    {\ifthenelse{\equal{\oneColFormat}{true}}{#1}{#2}}

\newtheorem{definition}{Definition}
\newtheorem{example}{Example}

\newcommand{\VS}{\textsf{VS}}
\newcommand{\FS}{\textsf{FS}}
\newcommand{\PS}{\textsf{PS}}

\newcommand{\TS}{\textsf{TS}}

\newcommand{\ar}[1]{\ensuremath{\textsf{ar}(#1)}}



%% file: statistics_checking.tex
\begin{table}{t}
\begin{footnotesize}
\begin{longtable}{||r|r|r|r|r|r|r|r|r||}
\hline\hline
\textbf{Ar\textbackslash Cnds}  & \textbf{1} & \textbf{1 (AVG)}& \textbf{2}& \textbf{2 (AVG)} & \textbf{3}& \textbf{3 (AVG)} & \textbf{4} & \textbf{4 (AVG)} \\ \hline\hline
1 (85 pr) &  19,064 &   224 &    53,530 &    630 &   180,246 &  2,121 &   298,292 &   3,509\\ \hline
2 (74 pr) & 110,092 & 1,488 &   207,871 &  2,809 &   221,061 &  2,987 &   477,440 &   6,452\\ \hline
3 (47 pr) & 294,962 & 6,276 & 3,757,208 & 79,941 & 3,806,917 & 80,998 & 6,127,015 & 130,362\\ \hline
4 (12 pr) &   5,116 &   426 &    12,939 &  1,078 &    22,508 &  1,876 &    30,300 &   2,525\\ \hline
%
%
\end{longtable}
\vspace*{-6mm}
\caption{Predicate query matching times ($\mu$s).\label{tab:ass_check_time}}
\end{footnotesize}
\vspace*{-3mm}
\end{table}

%% file: deepfind_appendix_common.tex

\section{Example code}
\label{sec:code}

\vfill 
Sample code found with \kbd{math\_graph} structure:
\vspace{-3mm}
\begin{figure}[!h]
  \centering
\prettylstformat
\begin{lstlisting}
:- module(named_graphs,	[complete_graph/2, cycle_graph/2], []).

:- use_module(library(lists), [append/3]).

complete_graph(N, graph(V,E)) :-
	count(N, V),
	generate_complete_edges(V, E).

generate_complete_edges(V, E) :-
	generate_complete_edges_(V, V, E).

generate_complete_edges_([], _, []).
generate_complete_edges_([V|Vs], AllV, E) :- 
	generate_complete_edges_for_vertex(V, AllV, E1),
	append(E1, RestE, E),
	generate_complete_edges_(Vs, AllV, RestE).

generate_complete_edges_for_vertex(_, [], []) :- !.
generate_complete_edges_for_vertex(V, [V|Vs], E) :- !,
	generate_complete_edges_for_vertex(V, Vs, E).
generate_complete_edges_for_vertex(V, [V1|Vs], [(V, V1)|E]) :-	
	generate_complete_edges_for_vertex(V, Vs, E).

cycle_graph(N, graph(V,E)) :-
	N = 2, !,
	V = [1,2],
	E = [(1,2),(2,1)].
cycle_graph(N, graph(V,E)) :-
	N > 1,
	count(N, V),
	generate_cycle_edges(V, E).

generate_cycle_edges([V1], [(V1, 1)]) :- !.
generate_cycle_edges([V1, V2|Vs], [(V1, V2)|Edges]) :-
	generate_cycle_edges([V2|Vs], Edges).

count(N, Lst) :-
	count_(1, N, Lst).
count_(I, N, []) :-
	I > N, !.
count_(I, N, [I|L]) :-
	I1 is I+1,
	count_(I1, N, L).
\end{lstlisting}
\vspace{-4mm} 
  \caption{\kbd{named\_graphs.pl} (Ciao library)}
  \label{fig:named_graphs}
\vspace{-8mm} 
\end{figure}
\vfill

\clearpage

\ \\
\vfill 
Sample code found with \kbd{al\_graph} structure:

\begin{figure}[!h]
  \centering
\prettylstformat
\begin{lstlisting}
:- module(ugraphs, [add_vertices/3], [assertions,isomodes] ).

:- use_module(library(sets), [ord_union/3]).
:- use_module(library(sort), [sort/2]).

:- pred add_vertices(+Graph1, +Vertices, -Graph2)
#  "Is true if @var{Graph2} is @var{Graph1} with @var{Vertices} added to it.".
add_vertices(Graph0, Vs0, Graph) :-
	sort(Vs0, Vs),
	Vs = Vs0,
	vertex_units(Vs, Graph1),
	graph_union(Graph0, Graph1, Graph).
% ...
\end{lstlisting}
  \caption{Fragment from \kbd{ugraphs.pl} (Ciao library).}
  \label{fig:ugraphs}
\end{figure}

\vfill

 
\clearpage
\input algorithms


\clearpage
\section{Additional tables}
\label{sec:anatables}

\input statistics_report_reduced

\input restore_data


%% file: algorithms.tex

\clearpage
\section{Algorithms for predicate matching}
\label{sec:algomatch}

The algorithms presented in this section are used to decide whether a
predicate is proven to match a condition (that condition is checked or
false) or that it cannot say anything about that property holding
(check).

\begin{algorithm}[!ht]
  \label{alg:matchcall}
  \caption{Matching Status of a calls condition for a predicate $p$}
  \textbf{Input:}
  $Analysis(P, \DD, \Q_\alpha)$, 
  $p \in exported(P)$,  
  $C = \calls(H, (Pre_1 ; ... ; Pre_n))$
  
  \textbf{Output:} Status of proof
  \begin{algorithmic}[1]
    \If{$\forall \tuple{H, \lambda^c, \lambda^s} \in
      Analysis\ s.t.\ H = p(X_1, ..., X_n),
      \bigvee_i\lambda^-_{TS(Pre_i, P)} \sqsupseteq \lambda^c$}
    \State{Status = \textbf{Checked}}
    \ElsIf{$\forall  \tuple{H, \lambda^c, \lambda^s} \in
      Analysis\ s.t.\ H =  p(X_1, ..., X_n),
       \bigvee_i\lambda^-_{TS(Pre_i, P)} \sqcap \lambda^c = \bot$}
    \State{Status = \textbf{False}}
    \Else
    \State{Status = \textbf{Check}}
    \EndIf
  \end{algorithmic}
\end{algorithm}

\begin{algorithm}[!ht]
  \label{alg:matchsuccess}
  \caption{Matching Status of a success condition for a predicate $p$}
  \textbf{Input:}
  $Analysis(P, \DD, \Q_\alpha)$, $p \in P$, $C = \success(H, Pre, Post)$
  
  \textbf{Output:} Status of proof
  \begin{algorithmic}[1]

    \If{$\exists\ \tuple{H, \lambda^c, \lambda^s} \in Analysis$
      s.t. $H = p(X_1, ..., X_n), \lambda^c = \lambda^+_{TS(Pre, P)}$}
    \If{$\lambda^s \sqsubseteq \lambda^-_{TS(Post, P)}$}
    \State{Status = \textbf{Checked}}
    \ElsIf{$\lambda^s \sqcap \lambda^+_{TS(Post, P)} = \bot$}
    \State{Status = \textbf{False}}
    \Else
    \State{Status = \textbf{Check}, analysis accurate enough}
    \EndIf
    \ElsIf{$\exists \ \tuple{H, \lambda^c, \lambda^s} \in Analysis$
      s.t. $H = p(X_1, ..., X_n), \lambda^c \sqsupset \lambda^+_{TS(Pre, P)}$}
    \If{$\lambda^s \sqsubseteq \lambda^-_{TS(Post, P)}$}
    \State{Status = \textbf{Checked}}
    \ElsIf{$\lambda^s \sqcap \lambda^+_{TS(Post, P)} = \bot$}
    \State{Status = \textbf{False}}
    \Else
    \State{Status = \textbf{Check}, Refine analysis}
    \EndIf
    \Else
    \State{Status = \textbf{Check}, No information for that calls, Refine analysis}
    \EndIf
  \end{algorithmic}
\end{algorithm}


%% file: statistics_report_reduced.tex
\begin{small}

\begin{longtable}{||p{32mm}|r|r|r|r|r|r||}
  \caption{Analysis statistics from core/lib modules: time($ms$) and memory($B$) consumption.\label{tab:statsreduced}}\\ \hline\hline
  \textbf{Module name}  &
  \multicolumn{1}{|p{10mm}|}{\textbf{load time}} &
  \multicolumn{1}{|p{13mm}|}{\textbf{regtype ana time}} &
  \multicolumn{1}{|p{18mm}|}{\textbf{regtype global stack mem}} &
  \multicolumn{1}{|p{10mm}|}{\textbf{shfr ana time}} &
  \multicolumn{1}{|p{17mm}|}{\textbf{shfr global stack mem}} &
  \multicolumn{1}{|p{10mm}|}{\textbf{total analysis time}} \\ \hline\hline
\endfirsthead
\caption{Analysis statistics from core/lib modules: time($ms$) and memory($B$) consumption. \emph{(continued)}.}\\ \hline\hline
  \textbf{Module name}  &
  \multicolumn{1}{|p{10mm}|}{\textbf{load time}} &
  \multicolumn{1}{|p{13mm}|}{\textbf{regtype ana time}} &
  \multicolumn{1}{|p{18mm}|}{\textbf{regtype global stack mem}} &
  \multicolumn{1}{|p{10mm}|}{\textbf{shfr ana time}} &
  \multicolumn{1}{|p{17mm}|}{\textbf{shfr global stack mem}} &
  \multicolumn{1}{|p{10mm}|}{\textbf{total analysis time}} \\ \hline\hline
\endhead
 \multicolumn{6}{c}{\emph{(continued in next page)}}\\
 \endfoot
 \endlastfoot
dict 	&	480	&	20	&	669,312	&	3,712	&	772,472	&	3,732	\\ \hline
sets 	&	548	&	116	&	1,462,696	&	1,512	&	1,923,720	&	1,628	\\ \hline
assrt\_write 	&	760	&	172	&	1,404,136	&	1,240	&	420,392	&	1,412	\\ \hline
sort 	&	544	&	184	&	877,104	&	992	&	222,288	&	1,176	\\ \hline
optparse\_tr 	&	744	&	32	&	814,168	&	1,068	&	950,272	&	1,100	\\ \hline
translation 	&	516	&	108	&	3,415,632	&	564	&	471,456	&	672	\\ \hline
exsteps 	&	664	&	28	&	990,872	&	296	&	1,186,552	&	324	\\ \hline
assrt\_write0 	&	724	&	80	&	1,030,808	&	96	&	271,688	&	176	\\ \hline
assrt\_lib\_extra 	&	724	&	108	&	1,103,072	&	48	&	314,680	&	156	\\ \hline
term\_list 	&	488	&	72	&	867,120	&	24	&	160,944	&	96	\\ \hline
civil\_registry 	&	508	&	76	&	554,032	&	16	&	621,504	&	92	\\ \hline
assertions\_props 	&	556	&	44	&	1,385,760	&	40	&	1,722,832	&	84	\\ \hline
pl2wam\_tables 	&	484	&	40	&	2,887,440	&	32	&	3,086,248	&	72	\\ \hline
embedded\_tr 	&	832	&	24	&	680,736	&	44	&	792,152	&	68	\\ \hline
terms 	&	528	&	44	&	571,912	&	12	&	653,248	&	56	\\ \hline
ceval1 	&	496	&	48	&	522,152	&	4	&	582,896	&	52	\\ \hline
unittest\_base 	&	516	&	36	&	630,600	&	16	&	740,344	&	52	\\ \hline
ceval2 	&	528	&	44	&	522,320	&	4	&	583,064	&	48	\\ \hline
errhandle 	&	540	&	28	&	620,024	&	16	&	747,456	&	44	\\ \hline
goal\_trans 	&	484	&	32	&	582,088	&	12	&	673,952	&	44	\\ \hline
llists 	&	480	&	24	&	526,384	&	12	&	592,568	&	36	\\ \hline
file\_utils 	&	564	&	20	&	641,728	&	12	&	758,024	&	32	\\ \hline
foreign\_compilation 	&	584	&	28	&	541,280	&	4	&	618,072	&	32	\\ \hline
qsort 	&	484	&	24	&	483,552	&	8	&	528,312	&	32	\\ \hline
srcdbg 	&	720	&	4	&	2,540,064	&	28	&	2,570,376	&	32	\\ \hline
meta\_props 	&	500	&	24	&	496,800	&	4	&	535,128	&	28	\\ \hline
strings 	&	532	&	16	&	693,304	&	12	&	809,928	&	28	\\ \hline
libpaths 	&	552	&	16	&	439,304	&	8	&	475,040	&	24	\\ \hline
metatypes\_tr 	&	468	&	24	&	439,216	&	0	&	477,136	&	24	\\ \hline
attr\_bench 	&	796	&	16	&	2,641,584	&	4	&	2,737,680	&	20	\\ \hline
between 	&	472	&	16	&	438,144	&	4	&	467,216	&	20	\\ \hline
iso\_char 	&	496	&	12	&	599,032	&	8	&	679,024	&	20	\\ \hline
length 	&	540	&	20	&	437,240	&	0	&	456,896	&	20	\\ \hline
phrase\_test 	&	512	&	8	&	556,144	&	8	&	644,392	&	16	\\ \hline
optparse\_rt 	&	488	&	4	&	456,144	&	8	&	501,440	&	12	\\ \hline
relationships 	&	532	&	8	&	479,552	&	4	&	519,952	&	12	\\ \hline
res\_exectime\_rt 	&	632	&	8	&	480,568	&	4	&	495,480	&	12	\\ \hline
resources\_tr 	&	476	&	8	&	419,248	&	4	&	444,992	&	12	\\ \hline
resources\_types 	&	484	&	8	&	483,864	&	4	&	534,696	&	12	\\ \hline
streams 	&	532	&	8	&	468,608	&	4	&	518,952	&	12	\\ \hline
ttyout 	&	500	&	8	&	450,688	&	4	&	498,456	&	12	\\ \hline
bundle\_params 	&	484	&	4	&	2,476,504	&	4	&	2,499,576	&	8	\\ \hline
ctrlcclean 	&	524	&	8	&	396,168	&	0	&	428,456	&	8	\\ \hline
miscprops 	&	460	&	4	&	448,888	&	4	&	480,056	&	8	\\ \hline
odd 	&	488	&	4	&	407,320	&	4	&	421,968	&	8	\\ \hline
old\_database 	&	492	&	4	&	494,040	&	4	&	529,896	&	8	\\ \hline
pretty\_names 	&	488	&	4	&	416,456	&	4	&	432,328	&	8	\\ \hline
dict\_types 	&	512	&	4	&	413,912	&	0	&	439,584	&	4	\\ \hline
fastrw 	&	512	&	0	&	447,968	&	4	&	471,416	&	4	\\ \hline
prf\_ticks\_rt 	&	636	&	0	&	506,008	&	4	&	520,416	&	4	\\ \hline
res\_nargs\_res 	&	524	&	0	&	393,080	&	4	&	407,560	&	4	\\ \hline
test1 	&	500	&	4	&	367,368	&	0	&	382,456	&	4	\\ \hline
test4 	&	520	&	4	&	372,968	&	0	&	384,120	&	4	\\ \hline
assrt\_synchk 	&	496	&	0	&	375,848	&	0	&	384,968	&	0	\\ \hline
c\_itf\_props 	&	480	&	0	&	414,208	&	0	&	435,624	&	0	\\ \hline
compressed\_bytecode 	&	500	&	0	&	367,288	&	0	&	376,488	&	0	\\ \hline
doc\_flags 	&	512	&	0	&	426,312	&	0	&	455,768	&	0	\\ \hline
doc\_props 	&	520	&	0	&	366,272	&	0	&	375,344	&	0	\\ \hline
regtypes\_tr 	&	484	&	0	&	415,608	&	0	&	434,712	&	0	\\ \hline
res\_litinfo 	&	528	&	0	&	498,792	&	0	&	526,104	&	0	\\ \hline
runtime\_ops\_tr 	&	460	&	0	&	375,136	&	0	&	389,048	&	0	\\ \hline
test2 	&	488	&	0	&	367,704	&	0	&	382,864	&	0	\\ \hline
unittest\_examples 	&	472	&	0	&	384,632	&	0	&	396,216	&	0	\\ \hline
TOTAL (63) & 34,088	& 1,680	& 47,436,912 & 9,924 & 44,316,888 & 11,604 \\ \hline
AVG & 541 & 26.7 & 752,967 & 157  & 703,443 & 184\\ \hline

\end{longtable}
\end{small}


%% file: restore_data.tex

\begin{small}

\begin{longtable}{||p{40mm}|r|r||}
\caption{Analysis dump files statistics from core/lib modules.\label{tab:statsrestored}}\\ \hline\hline
\textbf{Module name}  &
\multicolumn{1}{|p{16mm}|}{\textbf{dump size(B)}} &
\multicolumn{1}{|p{13mm}|}{\textbf{restore time(s)}} \\ \hline\hline
\endfirsthead
\caption{Analysis dump files statistics from core/lib modules.\label{tab:statsrestored}}\\ \hline\hline
\textbf{Module name}  &
\multicolumn{1}{|p{16mm}|}{\textbf{dump size(B)}} &
\multicolumn{1}{|p{13mm}|}{\textbf{restore time(s)}} \\ \hline\hline
\endhead
 \multicolumn{3}{c}{\emph{(continued in next page)}} \\
 \endfoot
 \endlastfoot
assrt\_write 	&	566,132	&	2,440	\\ \hline
sort 	&	524,490	&	1,772	\\ \hline
translation 	&	314,227	&	1,652	\\ \hline
assrt\_write0 	&	142,058	&	1,228	\\ \hline
assrt\_lib\_extra 	&	138,689	&	1,132	\\ \hline
assertions\_props 	&	142,057	&	1,084	\\ \hline
sets 	&	212,735	&	1,028	\\ \hline
exsteps 	&	257,632	&	916	\\ \hline
term\_list 	&	103,583	&	780	\\ \hline
errhandle 	&	51,222	&	640	\\ \hline
attr\_bench 	&	47,920	&	548	\\ \hline
terms 	&	59,107	&	536	\\ \hline
phrase\_test 	&	37,034	&	516	\\ \hline
file\_utils 	&	50,810	&	484	\\ \hline
embedded\_tr 	&	80,129	&	440	\\ \hline
strings 	&	36,279	&	400	\\ \hline
optparse\_tr 	&	136,929	&	384	\\ \hline
unittest\_base 	&	46,705	&	356	\\ \hline
civil\_registry 	&	30,588	&	328	\\ \hline
dict 	&	106,704	&	308	\\ \hline
iso\_char 	&	26,702	&	300	\\ \hline
foreign\_compilation 	&	23,623	&	276	\\ \hline
goal\_trans 	&	44,771	&	276	\\ \hline
llists 	&	22,500	&	260	\\ \hline
ceval2 	&	24,122	&	248	\\ \hline
ceval1 	&	21,995	&	232	\\ \hline
qsort 	&	17,867	&	200	\\ \hline
pl2wam\_tables 	&	17,649	&	184	\\ \hline
ttyout 	&	9,631	&	164	\\ \hline
streams 	&	12,383	&	160	\\ \hline
metatypes\_tr 	&	12,959	&	156	\\ \hline
meta\_props 	&	19,571	&	148	\\ \hline
libpaths 	&	11,676	&	124	\\ \hline
old\_database 	&	13,462	&	112	\\ \hline
dict\_types 	&	6,807	&	108	\\ \hline
relationships 	&	7,951	&	108	\\ \hline
between 	&	11,756	&	100	\\ \hline
fastrw 	&	6,754	&	100	\\ \hline
ctrlcclean 	&	7,474	&	96	\\ \hline
srcdbg 	&	31,936	&	92	\\ \hline
miscprops 	&	6,056	&	88	\\ \hline
doc\_flags 	&	4,678	&	84	\\ \hline
optparse\_rt 	&	8,713	&	84	\\ \hline
res\_litinfo 	&	6,662	&	80	\\ \hline
resources\_tr 	&	8,358	&	80	\\ \hline
bundle\_params 	&	6,528	&	72	\\ \hline
c\_itf\_props 	&	2,474	&	68	\\ \hline
resources\_types 	&	3,864	&	64	\\ \hline
length 	&	4,503	&	60	\\ \hline
test2 	&	1,519	&	52	\\ \hline
test1 	&	1,517	&	48	\\ \hline
odd 	&	2,498	&	44	\\ \hline
pretty\_names 	&	4,875	&	44	\\ \hline
prf\_ticks\_rt 	&	2,271	&	44	\\ \hline
res\_exectime\_rt 	&	2,676	&	44	\\ \hline
runtime\_ops\_tr 	&	3,204	&	40	\\ \hline
res\_nargs\_res 	&	2,646	&	36	\\ \hline
compressed\_bytecode 	&	782	&	32	\\ \hline
regtypes\_tr 	&	884	&	28	\\ \hline
test4 	&	218	&	24	\\ \hline
unittest\_examples 	&	58	&	24	\\ \hline
assrt\_synchk 	&	58	&	20	\\ \hline
doc\_props 	&	396	&	20	\\ \hline
TOTAL (63) &	3,512,057	&	21,596	\\ \hline
AVG	&	55,747	&	343	\\ \hline

\end{longtable}
\end{small}


%% file: deepfind.bbl
\begin{thebibliography}{}

\bibitem[\protect\citeauthoryear{Bruynooghe}{Bruynooghe}{1991}]{bruy91}
{\sc Bruynooghe, M.} 1991.
\newblock {A} {P}ractical {F}ramework for the {A}bstract {I}nterpretation of
  {L}ogic {P}rograms.
\newblock {\em Journal of Logic Programming\/}~{\em 10}, 91--124.

\bibitem[\protect\citeauthoryear{Cabeza and Hermenegildo}{Cabeza and
  Hermenegildo}{2000}]{ciao-modules-cl2000-short}
{\sc Cabeza, D.} {\sc and} {\sc Hermenegildo, M.} 2000.
\newblock {A} {N}ew {M}odule {S}ystem for {P}rolog.
\newblock In {\em International Conference CL 2000}. LNAI, vol. 1861.
  Springer-Verlag, 131--148.

\bibitem[\protect\citeauthoryear{Cousot and Cousot}{Cousot and
  Cousot}{1977}]{Cousot77-short}
{\sc Cousot, P.} {\sc and} {\sc Cousot, R.} 1977.
\newblock {A}bstract {I}nterpretation: a {U}nified {L}attice {M}odel for
  {S}tatic {A}nalysis of {P}rograms by {C}onstruction or {A}pproximation of
  {F}ixpoints.
\newblock In {\em Proc. of POPL'77}. ACM Press, 238--252.

\bibitem[\protect\citeauthoryear{Gallagher and de~Waal}{Gallagher and
  de~Waal}{1994}]{gallagher-types-iclp94-short}
{\sc Gallagher, J.} {\sc and} {\sc de~Waal, D.} 1994.
\newblock Fast and {P}recise {R}egular {A}pproximations of {L}ogic {P}rograms.
\newblock In {\em Proc. of ICLP'94}. MIT Press, 599--613.

\bibitem[\protect\citeauthoryear{Hermenegildo, Puebla, Bueno, and
  Lopez-Garcia}{Hermenegildo et~al\mbox{.}}{2005}]{ciaopp-sas03-journal-scp}
{\sc Hermenegildo, M.}, {\sc Puebla, G.}, {\sc Bueno, F.}, {\sc and} {\sc
  Lopez-Garcia, P.} 2005.
\newblock {I}ntegrated {P}rogram {D}ebugging, {V}erification, and
  {O}ptimization {U}sing {A}bstract {I}nterpretation (and {T}he {C}iao {S}ystem
  {P}reprocessor).
\newblock {\em Science of Computer Programming\/}~{\em 58,\/}~1--2 (October),
  115--140.

\bibitem[\protect\citeauthoryear{Hermenegildo, Bueno, Carro, L\'{o}pez, Mera,
  Morales, and Puebla}{Hermenegildo
  et~al\mbox{.}}{2012}]{hermenegildo11:ciao-design-tplp-short}
{\sc Hermenegildo, M.~V.}, {\sc Bueno, F.}, {\sc Carro, M.}, {\sc L\'{o}pez,
  P.}, {\sc Mera, E.}, {\sc Morales, J.}, {\sc and} {\sc Puebla, G.} 2012.
\newblock {A}n {O}verview of {C}iao and its {D}esign {P}hilosophy.
\newblock {\em TPLP\/}~{\em 12,\/}~1--2, 219--252.
\newblock http://arxiv.org/abs/1102.5497.

\bibitem[\protect\citeauthoryear{Maarek, Berry, and Kaiser}{Maarek
  et~al\mbox{.}}{1991}]{maarek1991information}
{\sc Maarek, Y.~S.}, {\sc Berry, D.~M.}, {\sc and} {\sc Kaiser, G.~E.} 1991.
\newblock An information retrieval approach for automatically constructing
  software libraries.
\newblock {\em Software Engineering, IEEE Transactions on\/}~{\em 17,\/}~8,
  800--813.

\bibitem[\protect\citeauthoryear{McMillan, Hariri, Poshyvanyk, Cleland-Huang,
  and Mobasher}{McMillan et~al\mbox{.}}{2012}]{mcmillan2012recommending}
{\sc McMillan, C.}, {\sc Hariri, N.}, {\sc Poshyvanyk, D.}, {\sc Cleland-Huang,
  J.}, {\sc and} {\sc Mobasher, B.} 2012.
\newblock Recommending source code for use in rapid software prototypes.
\newblock In {\em Proceedings of the 34th International Conference on Software
  Engineering}. IEEE Press, 848--858.

\bibitem[\protect\citeauthoryear{Mitchell}{Mitchell}{2008}]{mitchell:hoogle_19_nov_2008}
{\sc Mitchell, N.} 2008.
\newblock Hoogle overview.
\newblock {\em The Monad.Reader\/}~12 (November), 27--35.

\bibitem[\protect\citeauthoryear{Muthukumar and Hermenegildo}{Muthukumar and
  Hermenegildo}{1991}]{freeness-iclp91}
{\sc Muthukumar, K.} {\sc and} {\sc Hermenegildo, M.} 1991.
\newblock {C}ombined {D}etermination of {S}haring and {F}reeness of {P}rogram
  {V}ariables {T}hrough {A}bstract {I}nterpretation.
\newblock In {\em International Conference on Logic Programming (ICLP 1991)}.
  {MIT} {P}ress, 49--63.

\bibitem[\protect\citeauthoryear{Muthukumar and Hermenegildo}{Muthukumar and
  Hermenegildo}{1992}]{ai-jlp}
{\sc Muthukumar, K.} {\sc and} {\sc Hermenegildo, M.} 1992.
\newblock {C}ompile-time {D}erivation of {V}ariable {D}ependency {U}sing
  {A}bstract {I}nterpretation.
\newblock {\em Journal of Logic Programming\/}~{\em 13,\/}~2/3 (July),
  315--347.

\bibitem[\protect\citeauthoryear{Puebla, Bueno, and Hermenegildo}{Puebla
  et~al\mbox{.}}{2000a}]{assert-lang-disciplbook-short}
{\sc Puebla, G.}, {\sc Bueno, F.}, {\sc and} {\sc Hermenegildo, M.} 2000a.
\newblock {A}n {A}ssertion {L}anguage for {C}onstraint {L}ogic {P}rograms.
\newblock In {\em {A}nalysis and {V}isualization {T}ools for {C}onstraint
  {P}rogramming}. Number 1870 in LNCS. Springer-Verlag, 23--61.

\bibitem[\protect\citeauthoryear{Puebla, Bueno, and Hermenegildo}{Puebla
  et~al\mbox{.}}{2000b}]{assrt-theoret-framework-lopstr99}
{\sc Puebla, G.}, {\sc Bueno, F.}, {\sc and} {\sc Hermenegildo, M.} 2000b.
\newblock {C}ombined {S}tatic and {D}ynamic {A}ssertion-{B}ased {D}ebugging of
  {C}onstraint {L}ogic {P}rograms.
\newblock In {\em Logic-based Program Synthesis and Transformation
  (LOPSTR'99)}. Number 1817 in LNCS. Springer-Verlag, 273--292.

\bibitem[\protect\citeauthoryear{Puebla, Correas, Hermenegildo, Bueno,
  {Garc\'{\i}a de la Banda}, Marriott, and Stuckey}{Puebla
  et~al\mbox{.}}{2004}]{mod-an-lopstrbook-shortalt}
{\sc Puebla, G.}, {\sc Correas, J.}, {\sc Hermenegildo, M.}, {\sc Bueno, F.},
  {\sc {Garc\'{\i}a de la Banda}, M.}, {\sc Marriott, K.}, {\sc and} {\sc
  Stuckey, P.~J.} 2004.
\newblock {A} {G}eneric {F}ramework for {C}ontext-{S}ensitive {A}nalysis of
  {M}odular {P}rograms.
\newblock In {\em {P}rogram {D}evelopment in {C}omputational {L}ogic}. Number
  3049 in LNCS. Springer-Verlag, 234--261.

\bibitem[\protect\citeauthoryear{Puebla and Hermenegildo}{Puebla and
  Hermenegildo}{1999}]{spec-jlp}
{\sc Puebla, G.} {\sc and} {\sc Hermenegildo, M.} 1999.
\newblock {A}bstract {M}ultiple {S}pecialization and its {A}pplication to
  {P}rogram {P}arallelization.
\newblock {\em J.~of Logic Programming. Special Issue on Synthesis,
  Transformation and Analysis of Logic Programs\/}~{\em 41,\/}~2\&3 (November),
  279--316.

\bibitem[\protect\citeauthoryear{Reiss}{Reiss}{2009}]{reiss2009semantics}
{\sc Reiss, S.~P.} 2009.
\newblock Semantics-based code search.
\newblock In {\em Proceedings of the 31st International Conference on Software
  Engineering}. IEEE Computer Society, 243--253.

\bibitem[\protect\citeauthoryear{Rollins and Wing}{Rollins and
  Wing}{1991}]{rollins1991specifications}
{\sc Rollins, E.~J.} {\sc and} {\sc Wing, J.~M.} 1991.
\newblock Specifications as search keys for software libraries.
\newblock In {\em ICLP}. Citeseer, 173--187.

\bibitem[\protect\citeauthoryear{Stulova, Morales, and Hermenegildo}{Stulova
  et~al\mbox{.}}{2014}]{asrHO-ppdp2014}
{\sc Stulova, N.}, {\sc Morales, J.~F.}, {\sc and} {\sc Hermenegildo, M.~V.}
  2014.
\newblock {A}ssertion-based {D}ebugging of {H}igher-{O}rder {(C)LP} {P}rograms.
\newblock In {\em 16th Int'l. ACM SIGPLAN Symposium on Principles and Practice
  of Declarative Programming (PPDP'14)}. ACM Press.

\bibitem[\protect\citeauthoryear{Vaucheret and Bueno}{Vaucheret and
  Bueno}{2002}]{eterms-sas02-short}
{\sc Vaucheret, C.} {\sc and} {\sc Bueno, F.} 2002.
\newblock {M}ore {P}recise yet {E}fficient {T}ype {I}nference for {L}ogic
  {P}rograms.
\newblock In {\em SAS'02}. Number 2477 in LNCS. Springer, 102--116.

\end{thebibliography}
